\documentclass[printer]{aa}
\usepackage[varg]{txfonts}  
\usepackage{natbib}         
\bibpunct{(}{)}{;}{a}{}{,}  

\begin{document}

\title{Precision age indicators that exploit chemically peculiar stars}

\author{Guy Worthey}
\institute{Washington State University, 1245 Webster Hall, Pullman, WA, 99164-2814\\
\email{gworthey@wsu.edu}}

\date{Received 22/05/2015 / Accepted 13/07/2015}

\abstract{The integrated light of distant star clusters and galaxies
  can yield information on the stellar formation epochs, chemical abundance
  mixtures, and initial stellar mass functions, and therefore improve
  our understanding of galaxy evolution. } {We would like to find a way to
  improve the determination of galaxy star formation history from
  integrated light spectroscopy. } {Several classes of chemically
  peculiar (CP) stars arise during the course of normal evolution in
  single stars and noninteracting binary stars. An aging stellar
  population has periods of time in which CP stars contribute to the
  integrated light, and others in which the contributions fade. The
  HgMn stars, for example, occupy a narrow temperature range of 10500
  to 16000 K, which maps to a narrow range of ages. Wolf-Rayet stars,
  He-poor stars, Bp-Ap stars, Am-Fm stars, and C stars all become very
  common in a normal stellar population at various ages between zero
  and several Gyr, fading in and out in a way that is analogous to features
  used in stellar spectral classification. We examine population
  fractions and light fractions in order to assess the feasibility of
  using CP stars as age tracers.}  {Even though CP stars do not
  usually dominate in number, there are enough of them so that the CP
  spectral features are detectable in high-quality integrated spectra
  of young and intermediate age stellar populations. The new technique
  should be calibratable and useful.  Furthermore, using CP signatures
  as age dating tools sidesteps reliance on photometry that is
  susceptible to dust and Balmer features that are susceptible to
  nebular fill-in.}  {}


\maketitle

\titlerunning{Precision age indicators using CP stars}
\authorrunning{G. Worthey}

\section{Introduction}

Tracing stellar population age by integrated light is a mainstay of
galaxy evolution interpretation. Somewhat ambiguously, yet still of great value,
blue photometric colors generally indicate galaxies with ongoing star
formation, while red colors indicate galaxies that have not formed
significant numbers of stars in the last few hundred million years. A
technique that often gives higher precision is using Balmer feature
strengths in young populations \citep{2013ApJ...779..170L} or in old
ones \citep{2000AJ....119.1645T}. While these considerations yield a
mean, light-weighted age fairly readily, uncertainties remain. First,
since all Balmer features measure approximately the same thing, namely
stellar temperature, they are equivalent to one measureable quantity,
and a mean age is all one can realistically hope to achieve, which is
rather far from the ultimate goal of uncovering the star formation
history. There are also astrophysical degeneracies such as
age-metallicity degeneracy at old ages \citep{1994ApJS...95..107W} and
age-initial mass function degeneracy and age-interstellar medium
degeneracy at young ages \citep{2005AIPC..783..280L}.

The motivation for this paper is to highlight a new strategy for age
dating stellar populations using only high-quality integrated light
spectra. Since the method relies on chemically peculiar (CP) stars, a
description of the subtypes most likely to be useful is in order.

Wolf-Rayet (WR) stars are mass-losing upper main-sequence
stars. Because of their extreme winds and subsequent mass loss,
portions of the WR stars that have undergone nuclear processing are
exposed and they become CP and exhibit emission line
features in their spectra.  The notable bump in the $\lambda = 4640 -
4690$\AA\ region \citep{1991ApJ...377..115C} is seen in a
morphologically heterogeneous handful of starbursting galaxies known
as Wolf-Rayet galaxies \citep{1999A&AS..136...35S}. This class of
galaxies is already being modeled and studied
\citep{2014ApJS..212...14L}.

Moving to cooler classes of CP stars, in stars with slow rotation and
radiative atmospheres a combination of species-dependent gravitational
settling and radiation pressure can lead to heavy species levitating
to the top of the photosphere, causing strong alterations in the
observed spectrum \citep{1970ApJ...160..641M} that explain at least
some of the large numbers of peculiar spectral types among BAF stars
\citep{1969AJ.....74..375C}. We are interested in the numerically most
common subtypes. They are He-weak (CP4) stars that span spectral types
B2-B8, HgMn (CP3) stars that span spectral types B6-A0, Bp-Ap stars
(CP2) that span spectral types B6-F4, and Am-Fm stars (CP1) that span
spectral types A0-F4 \citep{1996Ap&SS.237...77S};  the
designations in parentheses are from
\citet{1974ARA&A..12..257P}. 

The diffusion (heavy species levitation) timescales are short compared
to the evolutionary timescales \citep{1976ApJ...210..447M}, and so it
is probably a combination of binarism, rotation, and magnetic fields
that control the levitation CP phenomenon
\citep{2014A&A...561A.147B}. The fraction of CP stars as a function of
age cannot at present be predicted ab initio. However, the stars are
all main sequence or near main sequence, and are therefore amenable to
precise empirical scrutiny.

It is known that the various subtypes lie in well-segregated
temperature regimes \citep{1993ASPC...44..577N,1996Ap&SS.237...77S},
which is a crucial aid to their use as extragalactic age
indicators. The levitating CP stars are common. HgMn stars account
for 15\% of B8 stars, and over 60\% of A6 stars are Am or Ap
\citep{1996Ap&SS.237...77S}. For stellar populations at the correct
ages, their optical-UV spectral features should be strongly influenced
by  chemical peculiarities.

Finally, coolest of all, C stars arise late in the lives of  1.5
-- 4 $M_\odot$ stars on the asymptotic giant branch where a third
dredge-up convects sufficient C to the surface to tip the number ratio
C/O $>$ 1 \citep{2008A&A...482..883M}. The mass range translates to population
ages between 0.5 and 5 Gyr for heavy element abundances typical of the
Magellanic Clouds. When C begins to dominate, the spectra of these
giants transform from being shaped primarily by H$_2$O and TiO
molecular features to CN, C$_2$, CH, and similar, while CO remains
strong. These stars are very cool \citep{2001A&A...369..178B}, but are
bolometrically the brightest stars in the population. A measure of the
ratio of TiO to C$_2$ features, for example, would readily give an
estimate of the C star fraction, and therefore additional age
information. Unlike WR or levitating CP stars, for which blue or UV
spectra are most revealing, the telltale spectra for C stars should be
in the red or infrared.

In this Letter, we  let the topic of WR stars stand as
sufficiently demonstrated, in the sense that it is clear that WR
features can be seen in the integrated spectra of some galaxies, and
that seeing such features provides additional leverage on the ages of
the youngest stars present in the observed galaxy. Section 2 discusses
the feasibility of HgMn stars as age indicators. Section 3
gauges the observable effect of C stars on intermediate age stellar
populations, and Sect. 4 briefly discusses the other CP classes. Section
5 extends the discussion to the future of this technique, outlining
the groundwork necessary to make it viable.

\section{HgMn Stars}

According to \citet{1996Ap&SS.237...77S}, HgMn stars span spectral
classes A0 through B6 and temperatures from 10500 to 16000 K. This
translates to main-sequence turnoff masses in a range from 3 to 4 $M_\odot$,
and that in turn translates to a narrow age range of 75 to 150 Myr at solar
metallicity \citep{1994A&AS..106..275B}. While they exist in the
stellar population, their number fraction apparently hovers near
10\% of B7, B8, and B9 stars \citep{1996Ap&SS.237...77S}.

\begin{figure}
\includegraphics[width=10cm]{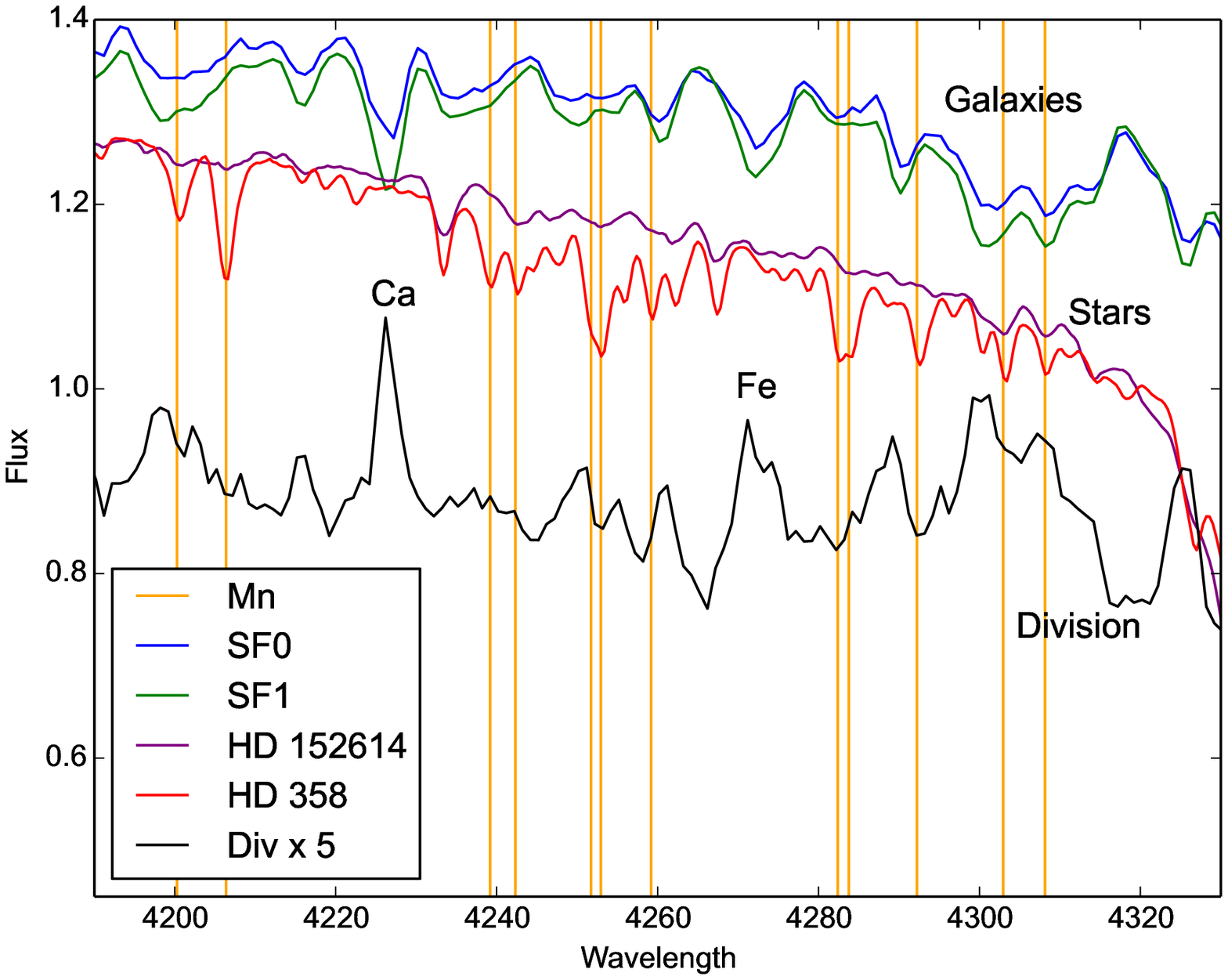}
\caption{Portion of the spectrum between H$\delta$ and
  H$\gamma$. Average galaxy spectra of categories SF0 (blue) and SF1
  (green) from \citet{2012MNRAS.420.1217D} and their division scaled
  about the average by a factor of five (black) are shown along with
  normal B8 star \object{HD 152614} ($\iota$ Oph; purple) and HgMn star \object{HD 358} ($\alpha$ And; red) from
  \citet{2004ApJS..152..251V}. Wavelengths of Mn transitions
  \citep{2013ApJS..205...14K} are marked as orange vertical lines. The
  feature of keenest interest is that the four strongest Mn features
  in HD 358 appear as small differential absorptions as one goes from
  SF1 to SF0 average galaxies.\label{fig1}}
\end{figure}

Their spectral classifications (and names) rest upon a \ion{Hg}{ii} feature at
$\lambda$3984 and \ion{Mn}{ii} features at $\lambda$3944, $\lambda$4137,
$\lambda$4206, and $\lambda$4282. There are a smattering of HgMn stars
in the Coude Feed Library \citep{2004ApJS..152..251V}, and we investigated
example stars to get a better idea of the true spectral signature for
this CP subclass, finding several more strong Mn features. An
illustrative pair of CP and normal stars is illustrated in Fig.
\ref{fig1}, slightly smoothed, along with wavelengths from
\citet{2013ApJS..205...14K}.

Also plotted in Fig. \ref{fig1} are two average blue galaxy spectra
from \citet{2012MNRAS.420.1217D}, their two bluest categories, SF0 and
SF1. The Mn features are not obvious until   the bluest
average (SF0) is divided  by the next (SF1), and then the four strongest Mn
features seen in the stars also show up as small extra differential
absorption. Most metallic features should be fading as the average
stellar population becomes hotter, as seen by the indicated Ca and Fe
features, and Mn is never strong except in HgMn stars,  so getting
more Mn absorption when moving to hotter stars is remarkable. In
combination with the fourfold wavelength coincidence, this is strong
evidence that the HgMn spectral signature has been detected in
aggregate galaxy spectra. It follows that the HgMn phenomenon can be
used in extragalactic age dating once appropriate models are
constructed.

\section{C stars}

The stellar evolutionary calculations of \citet{2008A&A...482..883M} include
the surface C/O abundance of AGB stars. We included them in the
\citet{1994ApJS...95..107W} population models, along with synthetic
C-rich spectra from
\citet{2009A&A...503..913A} where we imposed a correction for
self-extinction of $A_V = (2800 - T_{\rm eff})/800$ for stars with
$T_{\rm eff} < 2800$ K in order to compensate for the lack of dust in that
particular grid and make it come into agreement with the
\citet{2001A&A...369..178B} color temperature relations.

\begin{figure}
\includegraphics[width=10cm]{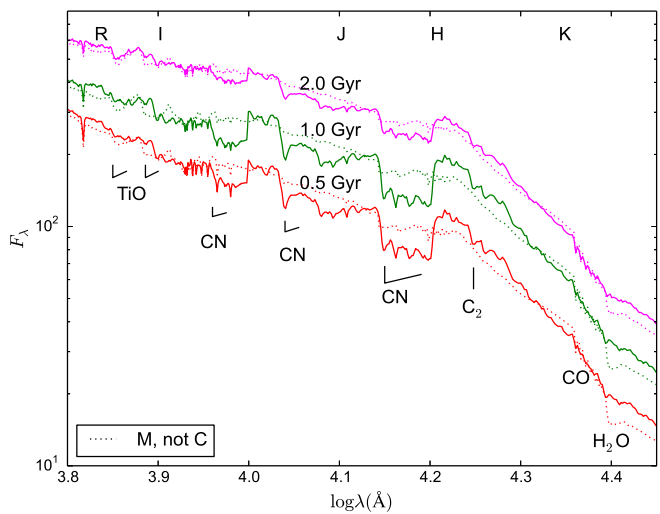}
\caption{Simple stellar population model spectra at solar abundance
  and three ages, 0.5, 1.0, and 2.0 Gyr, each age labeled on the
  plot. Carbon stars are included (solid lines) or M stars are
  substituted for C stars (dashed lines). While CO absorption stays
  about the same for both cases, there is marked and opposed variation in the
  H$_2$O and TiO features of M stars compared to C$_2$ and CN features
  due to molecular balancing. \label{fig2}}
\end{figure}

The low-resolution spectra from these calculations are displayed in
Fig. \ref{fig2} for simple stellar populations of solar abundance
and ages 0.5, 1.0, and 2.0 Gyr, and then displayed again with the C
stars replaced by M stars of the same atmospheric parameters. Redder
than $R$ band, the spectral changes are obvious.

For the purposes of this Letter, we do not adjust the models for recent
suggestions for downward revisions to the predicted number of C stars
\citep{2013ApJ...777..142G}; a glance at Fig. \ref{fig2} is
enough to understand that factors of two or three are not sufficient to
erase the C star spectra signatures, many of which exceed 25\% in that
figure. Ongoing discussion of the numbers, luminosity functions, and
frequencies of C stars are healthy steps toward using them as age
indicators.

One additional caution about C stars is that they are rare, evolved
stars susceptible to stochastic counting effects that can have effects
on integrated properties in some cases \citep{1997ApJ...479..764S}
though entire, large galaxies are probably immune
from stochastic effects. The other, bluer CP stars discussed in this
Letter are not susceptible, because they are main-sequence stars and are therefore
very numerous.

\section{Remaining CP Categories}

According to \citet{1993ASPC...44..577N}, who studied CP stars that
are members of open clusters, Si-strong and He-weak stars (also known
as hot CP2 stars or Bp-Ap stars) are already present at the 5\% level
in clusters of age 10 Myr, though they are not located at the main-sequence turnoff (MSTO) at that age, and they hover at 10\% frequency
thereafter until disappearing at around 500 Myr. The cooler CP2 stars
that show enhanced Sr, Cr, and Eu appear at $\sim$60 Myr and disappear
at $\sim$1 Gyr, never exceeding 5\% frequency. Am stars (CP1) also
appear at $\sim$60 Myr at the 5\% level (again, not at the MSTO), but
just as the hot CP2 stars are fading at 500 Myr, the Am frequency
appears to rise to 25\% before falling off again at or after 1
Gyr. North's study is limited especially at the old end by a dearth of
clusters.

\begin{figure}
\includegraphics[width=10cm]{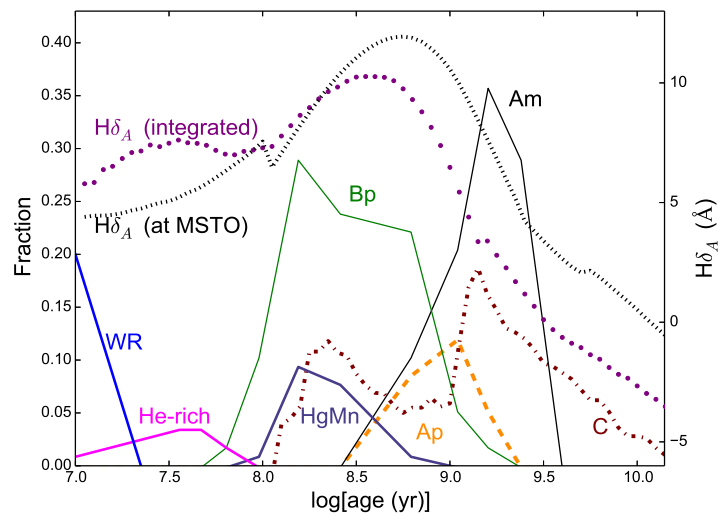}
\caption{Schematic time line  showing the age indicators. The H$\delta_A$ index
  \citep{1997ApJS..111..377W} for the hottest star near the main-sequence turnoff (MSTO; black broken line) and for the integrated
  light of the simple stellar population of solar abundance (purple
  dots) refer to the right-hand axis. The remainder of the lines refer
  schematically to the left-hand axis. The Wolf-Rayet fraction (blue
  line) is entirely schematic. The lines for He-rich (magenta line),
  Hg-Mn (slate blue line), Bp (thin green line), Ap (dashed orange
  line), and Am (thin black line) stars are estimated from data in
  \citet{1993ASPC...44..577N} with approximate stellar mass converted
  to a MSTO age and an overall normalization estimated
  from Bright Star Catalog classifications \citep{1996Ap&SS.237...77S}. The line
  for C stars (dash-dotted maroon line) is from the models, and therefore
  refers to the theoretical predictions of
  \citet{2008A&A...482..883M}, and is expressed as the fraction of C
  stars compared with the total number of AGB stars in the
  population. \citet{1996Ap&SS.237...77S} finds significantly
  more Am and HgMn stars and somewhat fewer
  Bp stars than \citet{1993ASPC...44..577N}. In this
  figure, the Bp star category includes both Si-strong and He-weak types, which
  transition at a boundary approximately where HgMn stars peak. There is
  probably a sharper boundary between Ap and Bp than is evident in
  this figure, and the old-age Am boundary is uncertain due to
  scarcity of data. \label{fig3}}
\end{figure}

These CP categories are shown in context with the others and with the
rise and fall of the Balmer feature index H$\delta_A$ in Figure
\ref{fig3}, which is an imprecise amalgamation of data and
models. If we can properly calibrate all the CP subtypes and then
detect their presence in integrated light, however, it follows that we
can use them as precise age dating tools.

\section{Discussion and conclusion}

Considering the comings and goings of hot and cool
CP2 stars, Am stars, HgMn stars, WR stars, and C stars, we see that
some event or other happens every few tenths along a log $t$ timeline
(Figure \ref{fig3}). This excellent ``age resolution''
is improved by including time-tested photometric colors and Balmer
features. Given this new wealth of observational signatures, we can
realistically consider deriving star formation histories from
integrated light alone.

The observability of the CP age indicators is good, though high S/N is
required for security. To estimate S/N requirements, we suppose that a
CP star has an increase (or decrease) in its various absorptions of
equivalent width $W_{CP}$. That equivalent width is attenuated by
stars in other parts of the HR diagram by a factor $f_{pop}$ and by
the numerical rarity of the CP star category $f_{CP}$ so that the
measured equivalent width is $W = W_{CP} f_{pop} f_{CP}$. The
uncertainty depends on S/N as $\sigma(W) = \sqrt{2} (\Delta\lambda -
W)/(S/N)$, where $\Delta\lambda$ is the window over which the
measurement is made, and the S/N ratio refers to that same spectral
window \citep{2006AN....327..862V}. For a single, weak absorption, if
a particular CP type is located at the main-sequence turnoff and has
10\% frequency ($f_{CP}=0.10$) and has a spectral feature that is 95\%
of continuum and about $\Delta\lambda =4$\AA\ wide ($W_{CP}\approx
4(0.05)=0.20$ \AA ), and if the turnoff stars contribute half the
light in the blue ($f_{pop}=0.50$), then $W \approx 0.01$\AA .  In
that case, one would want a continuum S/N $\sim 2000$ over the
wavelength span of the weak feature for secure detection. This
worst-case requirement is greatly lessened in the usual case where
many (if not hosts of) absorption features contribute to the signal,
and also many signature absorptions (especially H, He, and Ca) are
much stronger than 5\% deep.

The S/N requirement also lessens in the UV, where the absorptions are
stronger and the dilution by non-turnoff stars lessens. Working to
increase S/N requirements is that CP stars often have depressed
continuum in the UV, but it still works out to be a net positive gain
as long as the line absorption increases faster than the continuum
depression.

The observability as a function of CP type depends on the number and
strength of the absorption features in the observed spectral
window. Roughly speaking, however, relative to Hg-Mn stars, all the
types we mention will be easier to observe. He-rich stars are
characterized by a good collection of strong He lines throughout the
blue and visual, weak H lines, and all but absent C lines, the
combined equivalent widths of which should comfortably overwhelm the
Mn lines we point out for HgMn stars. Bp and Ap stars are the
more extreme cousins of Am stars, and show strong Si, Sr, Cr, and Eu,
again with a healthy sum of equivalent widths over the many features
present. Am stars have a weak Ca K feature combined with strong metal
lines, including Fe, and therefore again containing a very healthy sum
of total absorption. In the Am case, complete disambiguation of the Am
signature may require measuring Ca K accurately, worrisome because it
is a single spectral feature. To ease the worry, however, we note that
this is a strong absorption feature with $\Delta W \sim$1 to
2\AA\ when the Am peculiarity is imposed.

 We did not list some CP categories because of their relative
rarity. The small number of  CP stars lessens their influence on integrated
light. These include $\lambda$ Boo stars, metal-poor CH stars, and
stars that are CP because of binary evolution effects such as barium
stars and Sr4077 stars. These stars should be kept in mind for the
future, as their spectral signatures could, firstly, slightly blur the
results from the more common CP varieties, and, secondly and more
positively, some day enable a direct measurement of binary fractions in
external galaxies.

The calibration of the technique is in its infancy and needs much more
work. Spectral libraries should be built with explicit inclusion of
sequences of CP stars. If synthetic stellar libraries are employed, new grids
will have to be computed to accomodate the altered abundances, but
a more systematic study will be needed to specify the abundance patterns
that should be synthesized as representative of each class of CP
stars.

Spectral libraries should be extended to the UV, if possible, because
the spectral signatures of chemical peculiarity become stronger
\citep{1989A&AS...77..345F,1989A&AS...80..399F} and the light fraction
contributed by stars near the main-sequence turnoff increases, though
these trends are somewhat attenuated by a lowering of the UV continuum
in CP stars. Going to the UV is also convenient for application to
high-redshift galaxies.

For evolutionary population synthesis, the numbers and temperatures of
stars need to be known with confidence as a function of age and heavy element
abundance. In this regard, GAIA distances to thousands of nearby CP
stars will be very helpful  regarding temperature spreads and number
fractions, but continued survey work in open clusters, associations,
moving groups, and well-studied local group galaxies for CP stars is
even more essential because the ages and initial metallicities can be
known \citep{2007A&A...470..685L}.

In conclusion, this Letter presents an outline for a technique to
unravel age structures in the integrated light of stellar populations
using CP stars.

\begin{itemize}
\item{The CP technique has high age resolution compared to the Balmer
  feature technique because most CP stars turn on or
  turn off fairly suddenly as a function of population age.}
\item{The CP technique is insensitive to intervening interstellar dust.}
\item{The CP technique avoids the use of Balmer features, and therefore all
  complications arising from nebular emission.}
\item{The CP technique relies on high S/N galaxy spectra.}
\item{The CP technique also relies on
  a suitably thorough modeling effort that includes expanded stellar
  spectral libraries and better empirical constraints on number
  fractions of CP stars as a function of age, chemistry, and
  evolutionary state. }
\end{itemize}

\begin{acknowledgements}

I would like to thank C. R. Cowley for many stimulating conversations
regarding CP stars. I would like especially to thank S. C. Trager for
his encouragement and also my esteemed spectral library collaborators
P. Coelho, J. Falc{\'o}n-Barroso, S. R. Heap, A. Lan{\c c}on,
D. Lindler, L. P. Martins, R. F. Peletier, P. Prugniel,
P. S{\'a}nchez-Bl{\'a}zquez, D. Silva, and A. Vazdekis. The comments
of the anonymous referee were very helpful in improving this
Letter. Support for program AR-13900 was provided by NASA through a
grant from the Space Telescope Science Institute, which is operated by
the Association of Universities for Research in Astronomy, Inc., under
NASA contract NAS 5-26555.

\end{acknowledgements}

\bibliographystyle{aa}
\bibliography{ms}

\end{document}